\documentclass[prl, twocolumn, superscriptaddress, floatfix, tighten]{revtex4-2}

\usepackage{xcolor}
\usepackage{amsmath}
\usepackage{amssymb}
\usepackage{bm}      	
\usepackage{graphicx}
\usepackage[unicode=true,pdfusetitle,
 bookmarks=false,bookmarksnumbered=false,bookmarksopen=false,
 breaklinks=false,pdfborder={0 0 1},backref=false,colorlinks=true,urlcolor=blue,citecolor=blue,linkcolor=black]
 {hyperref}
\usepackage{subfigure}
\usepackage{textcomp}
\usepackage{microtype} 

\newcommand{\blue}[1]{{\color{blue}{#1}}}

\newcommand{\bsub}{\begin{subequations}}
\newcommand{\esub}{\end{subequations}}

\newcommand{\pup}[1]{{\scriptscriptstyle{({#1})}}}
\newcommand{\chalk}{\Upsilon}

\newcommand{\MM}[1]{\blue{#1}}

\begin{document}

\title
{
Fractal Confinement and Magnetic Self-sabotage of Current Flow:\\ Electrons Near the Metal-insulator Crossover in a 2D $\delta$-layer
}

\author{Xinghai Zhang}
\affiliation{Department of Physics, KTH Royal Institute of Technology, SE-106 91 Stockholm, Sweden}
\author{Matthew S. Foster}
\affiliation{Department of Physics and Astronomy, Rice University, Houston, Texas 77005, USA}
\author{Markus M\"uller}
\affiliation{PSI Center for Scientific Computing, Theory and Data, CH-5232 Villigen PSI, Switzerland}

\date{\today}

\begin{abstract}
We study low-$T$ transport and itinerant magnetism in 2D $\delta$-doped semiconductors. Large-scale Hartree-Fock calculations with random hoppings and repulsion $U_0$ reveal two transport regimes near a critical density. The conductivity $\sigma(T)$ initially increases as $T$ drops below $U_0 / 10$. However, at $T \sim U_0/100$, a tiny density of magnetic moments forms, driving an insulating downturn. Fractal wavefunctions locally amplify magnetic interactions, initially concentrating currents into branched filaments, 
but ultimately 
choking them off via dynamic SU(2) symmetry breaking.     
\end{abstract}
\maketitle

\emph{Introduction}---Semiconductors hosting dopants at precisely defined depths and lateral positions hold the promise of a future quantum computing  platform based on Rydberg gates~\cite{Kane1998, He2019, Edlbauer2025}. The recent interest in these systems has driven a spectacular improvement in the fabrication of ultra-thin “$\delta$-layers” of dopants~\cite{Constantinou2023, danna2025}.
The availability of these 2D sheets reignites the 40-year-old, yet still unsolved question: Does a genuine metallic phase of electrons exist in 2D? While disorder-induced Anderson localization rules this out for non-interacting particles with weak spin-orbit interaction \cite{Lee1985,Evers2008}, the fate of {\em interacting} electrons has remained a long-debated issue
\cite{BK1994,DS2005, Punn2001, Punn2005, Ani2007, dobrosavljevic2012conductor, DS2013, DS2014}. 
The development in $\delta$-layers is particularly interesting for two reasons: They display the metal-to-insulator (MI) crossover at much higher densities than in the more commonly studied electron gases induced at interfaces, accessing a different regime of competition between interactions and disorder. Moreover, the advent of ever higher resolution nanomagnetometry~\cite{Casola2018,Marchiori2022} 
enables increasingly {\em local} probes of the emerging magnetism and inhomogeneous current flow expected in shallow metallic $\delta$-layers at the brink of insulating breakdown. This promises unprecedented insight into 
intriguingly complex 
{physics,}
where the localization of quantum particles, the onset of magnetism \cite{Shashkin2001, Shashkin2006, Ani2006} and glassy 
features~\cite{Bogdanovich2002, Ovadyahu2013, Dobrosavljevic2012}, 
as well as the breakdown of screening~\cite{BK1994,Amini2014} go hand in hand.

Finkel'stein~\cite{Finkelstein1983} presented a 
scaling (RG) theory of the MI transition in bulk semiconductors subject to disorder and interactions. 
His unexpected result unfortunately limited its own predictive power: The conductance is \emph{enhanced} by Altshuler-Aronov 
(AA) 
\cite{AA1985}
corrections in the attractive triplet (ferromagnetic exchange) channel~\footnote{
By contrast, Coulomb interactions destroy the antilocalization predicted for non-interacting electrons with strong spin-orbit coupling,
due to the singlet-channel Altshuler-Aronov correction \cite{BK1994}.},
but the triplet coupling displays a runaway RG flow at a finite length or temperature scale. While this suggests a kind of Stoner instability of itinerant ferromagnetism, the runaway to strong coupling 
prevents the prediction of zero-temperature state properties \cite{BK1994}.

To understand the onset of magnetism in a 3D metal close to the MI transition, Bhatt and coworkers~\cite{Milo1989,Bhatt1992}
employed a Hartree-Fock (HF) treatment and found the paramagnetic metal to become unstable to the formation of local moments---patches of ferromagnetic magnetization in regions far more localized than the HF orbitals themselves, while the MI transition occurs at yet stronger disorder.
Here, using a similar approach, we revisit the 2D problem in the case of $\delta$-layers of randomly (nearly Poissonian) distributed, half-filled donor sites. The results turn out to differ significantly from the 3D case, shedding light on the low-temperature phase 
that evades Finkelstein’s theory, and the unexpected way a 2D metal becomes an insulator. Describing the system as a Hubbard model with inhomogeneous, distance-dependent hoppings $t_{ij}$ and an onsite repulsion $U_0$ evaluated for hydrogenic donor wavefunctions, our only tuning parameter is the 
dopant density
$ n =  2/ \sqrt{3} R^2$, 
where $R$ is the average inter-electron separation.
We find the MI crossover to occur close to $R= 2 a_0$, 
where $a_0$ is the donor Bohr radius\MM{,} and
$U_0/t_{\rm typ}\approx 0.89$
($t_{\rm typ}$ is the typical hopping strength) 
is substantially lower than $U_0/t\approx 8.5$ in Hubbard models on a triangular lattice without positional disorder~\cite{Szasz2020}.

Our key observations 
for a system close to the crossover 
are: 
{\em (i)} For $T \gtrsim U_0/100$, the conductivity $\sigma(T)$ increases logarithmically with decreasing $T$ and is accompanied by the increase of the density of states $\nu$ at the Fermi level; 
we attribute both to triplet AA corrections
\footnote{
While our HF calculations are sensitive to weak localization effects, they do not result in a  temperature dependence, since inelastic dephasing \cite{Aleiner1999Interaction,ChakravartySchmid1986} is absent in HF theory.
}. 
{\em (ii)} At $T \sim U_0/100$ local moments start to form spontaneously. Even though their spatial density remains tiny 
{($\sim 0.1\%$)}, 
we find a concomitant and significant downturn in $\sigma(T)$. 
Our HF analysis provides spatially-resolved maps of the strongly inhomogeneous  magnetization and conductivity (reflecting the multifractality of the HF wavefunctions \cite{Evers2008,Amini2014,Zhang2024}), unveiling an unexpected route to the insulator.  
                        
We confirm the old field-theory picture to be qualitatively correct up to the formation of moments. The concentration of HF orbitals (close to $E_F$) on a fractal support entails enhanced interactions \cite{Feigelman2007,Feigelman2010}.
In the attractive triplet channel, this first boosts the conductivity \cite{BK1994}, but when (locally) a Stoner instability is reached, moments form. 
We find that this happens precisely in places where the current is most enhanced. Thus, even few moments have a disproportionally large effect on the global current response, reversing the metallic $T$-dependence of $\sigma(T)$ to an insulating trend. 

We link the downturn in the conductivity with decreasing temperature to the turn-on of spin-flip scattering between local moments and conduction electrons; this breaks the spin symmetry of the latter dynamically and amputates the delocalizing effect of the triplet AA mechanism. 
The remaining AA effects in the density channel strongly suppress transport. 
At yet lower temperatures moments are expected 
to become correlated in clusters and eventually undergo a genuine time-reversal-symmetry breaking, presumably forming a spin glass. 
We conjecture the 
ultimate flow to a magnetic insulator at $T=0$ to be generic, but to set in at lower and lower temperatures in increasingly dense systems.  
The emerging local moments are 
formed by many overlapping critical states, not 
by single, anomalously localized states on particularly isolated sites. 
This results in an interesting situation where 
localized (magnetic moments) and extended (multifractal wavefunctions) 
features co-exist in a surprising way.

\begin{figure}[tb!]
    \centering
    \includegraphics[width=0.95\linewidth]{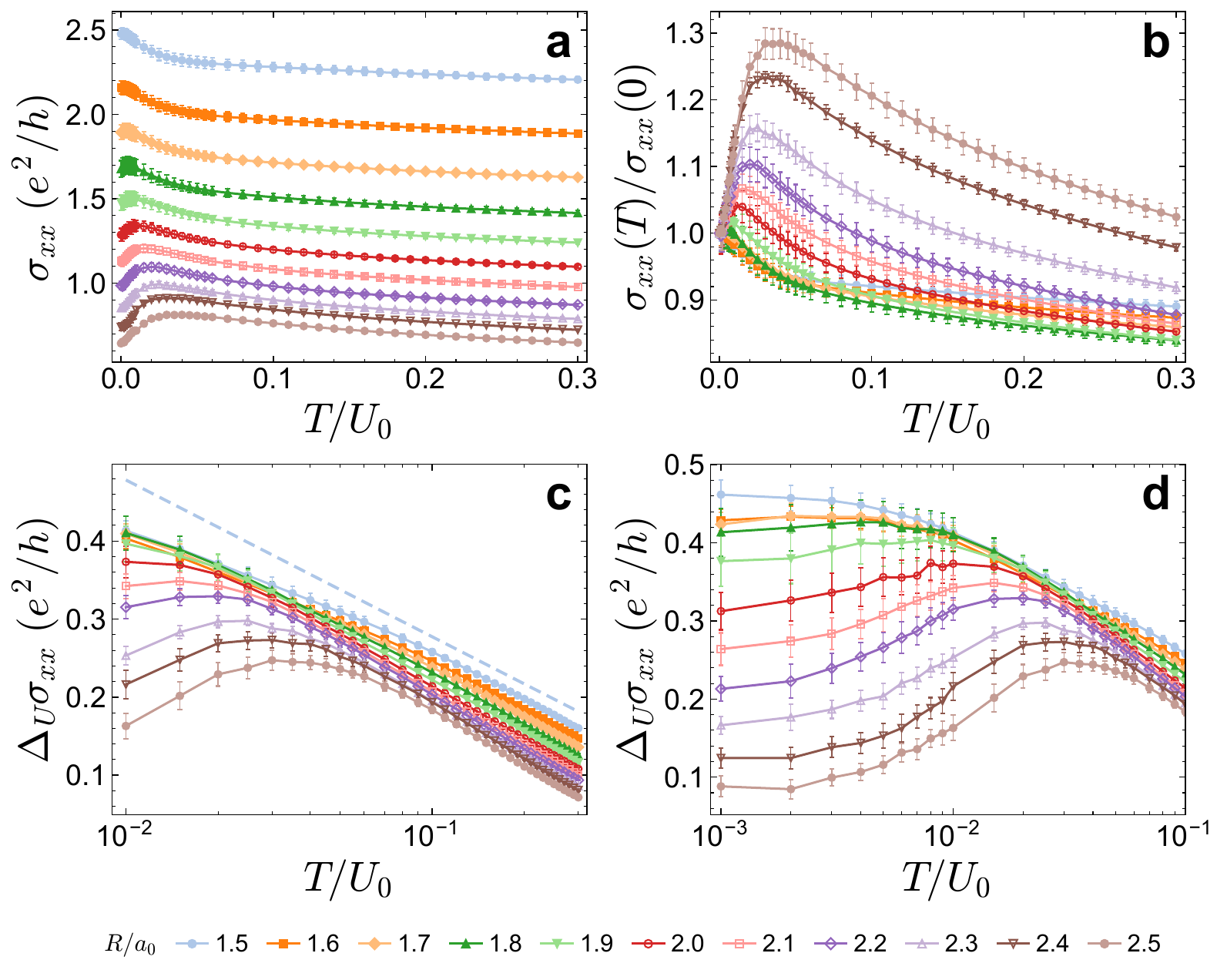}
    \caption{Longitudinal conductivity $\sigma_{xx}$ in distorted triangular lattices with 
    average lattice constant $R$.
    (a) $T$-dependence. 
    (b) $\sigma_{xx}$ scaled by its $T=0$ limit. 
    (c) The difference between interacting and non-interacting systems, 
    $\Delta_U \sigma_{xx}(T) \equiv \sigma_{xx}(U_0, T) - \sigma_{xx}(U=0, T)$, 
    displays a predominant 
    \emph{antilocalizing}
    spin-triplet Altshuler-Aronov correction [$ -\log(T)$] at high $T$. 
    (d) At lower $T$, $\Delta_U \sigma_{xx}(T)$ adopts a localizing $\log({T})$ trend, 
    most prominently for lower densities (larger $R$), 
    implying the dominance of the repulsive spin-singlet channel. 
    The saturation for the lowest temperatures is a finite-size effect.
    At $T=0$, the local magnetic moments
    break both spin SU(2) and time-reversal symmetries, 
    so the system localizes in the unitary class.
    Data is for a linear size $L = 100$, energy broadening $\eta = 0.01 U_0$ in the Kubo formula, 
    and results are averaged over $10$ disorder realizations at each $R$.} 
    \label{fig:sigmaT}
\end{figure}

\begin{figure*}
    \centering
    \includegraphics[width=0.8\linewidth]{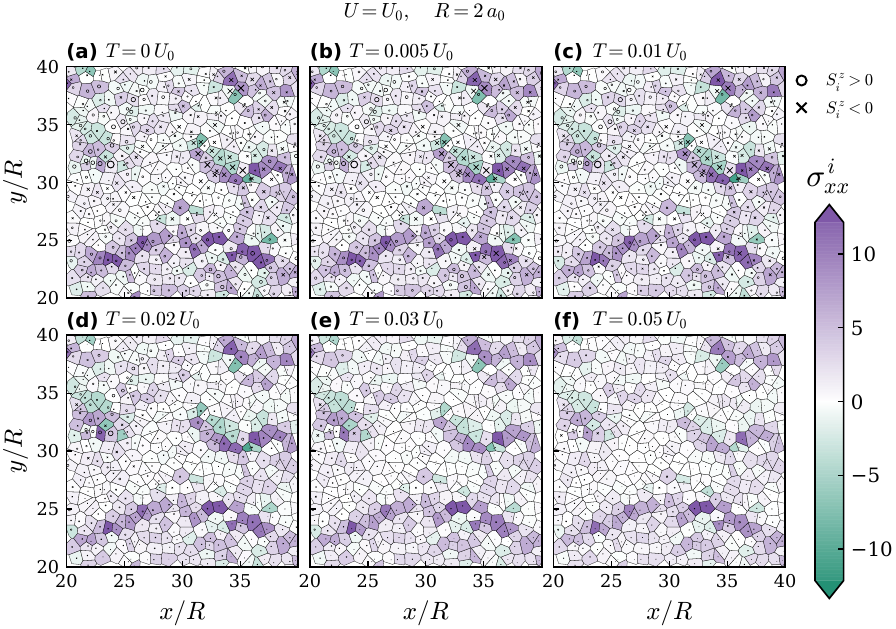}
    \caption{The local conductivity and local moments of a representative $20\times 20$ sub-area of a critically dense  ($R=2a_0$) sample of size $100\times 100$, at different temperatures. 
    The local longitudinal conductivity is represented by the filling color in the Wigner-Seitz  cell surrounding a site, while the magnitude of the local moments is indicated by the size (area proportional to ${|S_i^z|}$) of the circle (spin up) or cross (spin down). 
    Note the strong correlation between strong current flow and large polarization at low $T$.
    }
    \label{fig:sigmai}
\end{figure*}

\begin{figure*}
    \centering
    \includegraphics[width=0.8\linewidth]{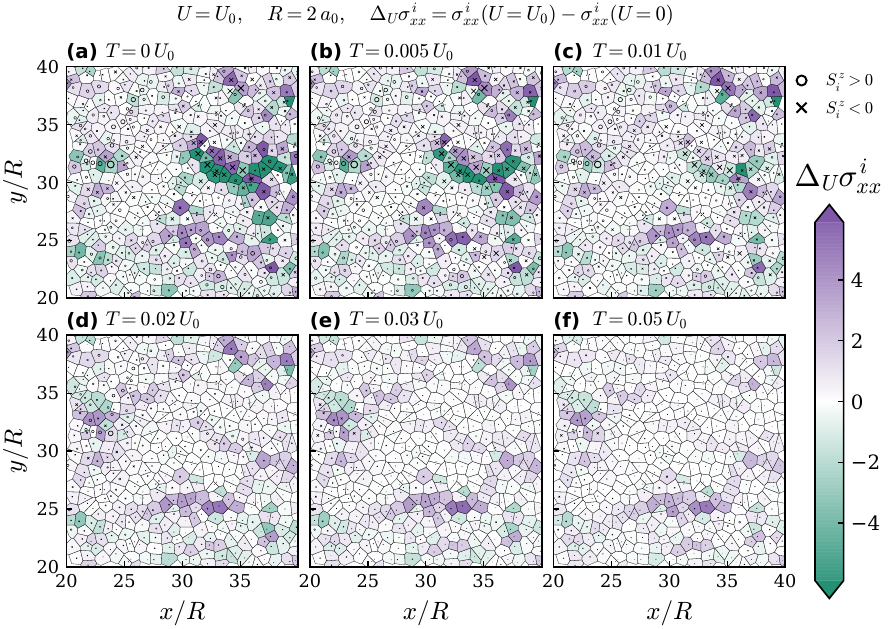}
    \caption{Local moments and the interaction-induced difference in local conductivity (as compared to a non-interacting system) in the same spatial region of the same sample as shown in Fig.~\ref{fig:sigmai}. }
    \label{fig:dsigmai}
\end{figure*}


\emph{Model and Setup}---We model the electrons in the $\delta$-layer by a two-dimensional tight-binding model of randomly distributed dopants hosting hydrogen-like orbitals,
\begin{equation}
    H = - \sum_{ij} t_{ij} c_{i\sigma}^\dagger c_{j\sigma} + U \sum_i n_{i\uparrow} n_{i\downarrow} - \mu \sum_i n_i\,.
    \label{eq:H}
\end{equation}
Here the hopping $t_{ij}=t_{ji}^*$ (real in the absence of an out-of-plane field) depends solely on the distance $r_{ij}$ between the  dopants $i,j$
\cite{SM}, and we only retain the local part of the Coulomb repulsion
with $U=U_0=5e^2/8\kappa a_0$, $\kappa$ being the host dielectric constant. 
As each dopant contributes one conduction electron, the chemical potential $\mu$ is  tuned to ensure half-filling.
We approximate a Poisson distributed ensemble of sites by a locally deformed triangular lattice with lattice constant $R$.  
Each site is sampled uniformly from a disk of radius $0.5R$ centered at the regular lattice position. In this Letter, we focus on the MI crossover regime near $R\approx 2a_0$. In silicon this corresponds to a dopant density of $n\approx 6.5\times 10^{12}$ cm$^{-2}$,
using an effective isotropic Bohr radius $a_0\approx 2.1$ nm, close to where the crossover is observed in experiments~\cite{Goh2006,danna2025,Mottnote1}.

The interaction term in Eq.~\eqref{eq:H} can be decoupled by a Hartree-Fock mean-field ansatz,
\begin{equation}
    n_{i\uparrow}n_{i\downarrow}=\frac{1}{2}\rho_{i}n_{i}-\mathbf{m}_{i}\cdot c_{i}^{\dagger}\boldsymbol{\sigma}c_{i}-\frac{1}{4}\rho_{i}^{2}+\mathbf{m}_{i}^{2}\,,
    \label{eq:HF}
\end{equation}
where the local density and spin polarization are 
$\rho_{i} =\left\langle n_{i}\right\rangle$ 
and 
$\mathbf{m}_{i}=\frac{1}{2}\langle c_{i}^{\dagger}\boldsymbol{\sigma}c_{i}\rangle$, 
respectively, evaluated as a thermal average over the occupations of HF orbitals. 
For simplicity, we assume that the local moments align along a preferred axis, which we identify as the $z$-axis in spin space, writing 
$\mathbf{m}_{i}=m_{i}^{z} {\bf \hat{z}}$~\footnote{Rashba-type SO couplings in the delta-layer might favor an out-of-plane orientation of the spins. Within local
clusters ferromagnetic correlations ensure alignment, while more distant clusters are only weakly correlated and polarize independently of each other, except at very low temperatures.}. 
The density and magnetization profiles are determined self-consistently 
\footnote{
Once local moments form,  
the HF has multiple solutions with similar local moment densities and global conductivity \cite{SM}. 
We present the physical quantities evaluated from one of the converged solutions, unless specified otherwise.
}. 
The global and local conductivity (current response to a uniform electric field) are subsequently evaluated using linear response theory.
By correlating the local transport coefficients with the spin polarization profile, we can determine the relationship between local interaction effects and spatial patterns of magnetism and transport.


\emph{Global transport}---The longitudinal conductivity, i.e. the total current response to an electric field, is evaluated using the Kubo-Greenwood formula. At finite $T$, one has
\begin{equation}
    \sigma_{xx}\left(\mu,T\right)
    =
    \int d\varepsilon\left(
    -\partial f/\partial \varepsilon\right)
    \sigma_{xx}\left(\varepsilon\right)\,,
    \label{eq:sigmaT}
\end{equation}
where $f=\left[e^{\beta\left(\varepsilon-\mu\right)}+1\right]^{-1}$ is the
Fermi-Dirac distribution function and $\sigma_{xx}(\varepsilon)$ is the 
conductivity of converged HF eigenstates near energy $\varepsilon$ 
\cite{SM}.

Fig.~\ref{fig:sigmaT} shows $\sigma_{xx}$ 
over a range of dopant densities and temperatures. The interaction has a significant effect on transport, displaying logarithmically enhanced conductivity as a result of Altshuler-Aronov (AA) corrections, 
which are dominated by the attractive triplet channel
{[see Eq.~(\ref{betasigma})]}. 
With varying density 
the conductivity exhibits distinct low-temperature behaviors, with an {\em apparent} separatrix near $R= 2 a_0$. 
At low densities ($R\geq 2a_0$) $\sigma_{xx}(T)$ reaches a maximum at some $T^*_\sigma(R)$
and then decreases as $T\to 0$. The crossover temperature  $T^*_\sigma(R)$ rises fairly rapidly with $R$, as one expects from Finkel'stein's runaway flow toward a magnetic instability (cf.\ End Matter).
At high dopant densities (corresponding to weak effective interactions) the conductivity $\sigma_{xx}(T)$ continues to increase slightly as the temperature is lowered. The absence of a maximum here is likely a finite-size effect.

The temperature-dependence becomes yet clearer when considering the difference in conductivity between the interacting and non-interacting systems, $\Delta_U \sigma_{xx}$, as shown in Figs.~\ref{fig:sigmaT}(c,d). 
In the high-$T$ regime, the interactions induce AA corrections which enhance the conductivity by an amount proportional to $-\log(T)$, with a prefactor that is nearly independent of the electron density (close to $R = 2 a_0$). 
However, as $T$ is further lowered,  $\Delta_U \sigma_{xx}(T)$ turns around and follows a logarithmic law  with the opposite sign of the prefactor.
Interestingly this MI crossover correlates with
the downturn in the density of states at the Fermi level 
and the emergence of local magnetic moments (see Fig.~\ref{fig:Lscaling}).
As a result of the reduction of spin SU(2) symmetry to U(1),
the spin-triplet AA correction is diminished to one-third of its original value and becomes subdominant compared to the localizing spin-singlet AA correction \cite{BK1994}. 
Interactions thus fail to stabilize the metallic phase at very low temperatures, where local moments spoil the conductivity enhancement. Surprisingly, the density of the local moments remains very low, a fraction of order 
$10^{-3}$ of all sites being magnetized at 
$T^*_\sigma$
and down to $T=0$.


\emph{Correlation of local current flow and magnetization}---
To understand how such a sparse set of 
moments can nonetheless be the cause of the  change from metallic to insulating $T$-dependence of the conductivity,
we numerically examine the spatially resolved current flow and spin polarization.
The local conductivity describes the local current density in response to a uniform 
electric field, $j{^i_\alpha} = \sum_\beta \sigma_{\alpha\beta}^i E_\beta$. 
It is evaluated via the following Kubo-Bastin-type formula, 
\begin{equation}
    \!\!
    {\sigma_{\alpha\beta}^i}
     = i {e^2}\int d\varepsilon\, f(\varepsilon)\, \mathsf{Tr}\!\left[
    \begin{aligned}
        & j^i_\alpha G_R'(\varepsilon) \mathcal{J}_\beta  \delta(\varepsilon-H) \\
        &-  j^i_\alpha \delta(\varepsilon-H) \mathcal{J}_\beta G_A'(\varepsilon)
    \end{aligned}
    \right],
    \!\!
    \label{eq:sigma-i}
\end{equation}
where the local paramagnetic current density is defined as
\begin{equation}
\label{J_i}
    \mathbf{j}_i
    = 
    \frac{1}{2\mathcal{A}_i}
    \sum_j 
    \left[
    -i(\mathbf{r}_i - \mathbf{r}_j) t_{ij} c^\dagger_i c_j + \textrm{H.c.} 
    \right]\,
\end{equation}
and $\mathcal{J}_\alpha =\sum_i j^i_\alpha \mathcal{A}_i$ is the volume integral of the current density.
Here $\mathcal{A}_i$ is the area 
of the Wigner-Seitz cell surrounding the site. 
Eq.~(\ref{J_i}) represents an areal average over the currents flowing through the bonds connecting to site $i$. 

Fig.~\ref{fig:sigmai} shows the formation of local magnetic moments alongside 
the evolution of 
${\sigma_{\alpha\beta}^i}$ 
as the temperature is lowered in systems 
with 
$R=2 a_0$. At high 
$T = 0.05 U_0$
[Fig.~\ref{fig:sigmai}(f)], there are no magnetic moments. The local conductivity is inhomogeneous because of the random dopant positions and concomitant modulated hopping. As $T$ 
decreases [Figs.~\ref{fig:sigmai}(c)--(e)], local moments form in regions where the current flow tends to be particularly strong. This is accompanied by a significant change of 
${\sigma_{\alpha\beta}^i}$
in their vicinity. At very low 
$T$ 
[Fig.~\ref{fig:sigmai}(a) and (b)], the local moments form ferromagnetically correlated clusters, as expected from crossing a local Stoner instability. 
Not only is the magnitude of 
${\sigma_{\alpha\beta}^i}$ 
much larger where moment clusters form, but also the current flow develops significant vorticity at low temperature, 
with the sign of the local conductivity fluctuating from site to site. 
Magnetic moments
are thus seen to significantly modify the local transport, inducing 
loop currents that do not contribute to the global conductivity, but rather degrade it.

To clarify the role of interactions in local-moment formation and  transport, we show the interaction-induced correction to the local conductivity
relative to the noninteracting system, $\Delta_U \sigma_{xx}^i$,
in Fig.~\ref{fig:dsigmai}. At high temperatures 
[Figs.~\ref{fig:dsigmai}(d)--(e)], the corrections are positive at most lattice sites, reflecting the dominance of the triplet-channel AA corrections. At lower temperatures where local moments form [Figs.~\ref{fig:dsigmai}(a)--(c)], the AA enhancement of the local conductivity persists at sites with weak or no magnetization. However, the magnitude of the conductivity corrections is much larger at and around clusters of local moments and can even be strongly negative. Indeed, when averaged over such clusters the local conductivity is found to diminish, rendering the system overall less conductive in this local-moment regime 
[$T\lesssim T_s^* \approx 0.015 \, U_0$ for $R=2a_0$, Fig.~\ref{fig:Lscaling}(a)]. 
In the $T\to 0$ limit we expect the system to become entirely insulating, $\sigma_{xx}(T\to 0)\to 0$. 

\begin{figure}
    \centering
    \includegraphics[width=0.95\linewidth]{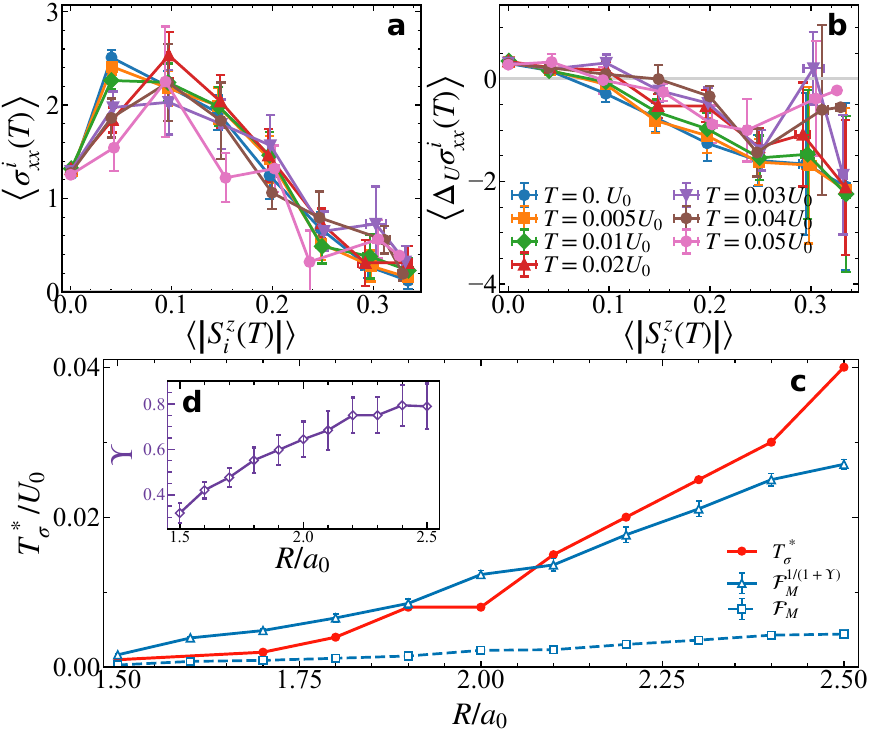}
    \caption{
    Correlation between local conductivity and local moments. $\sigma_{xx}^i$ and the interaction corrections to the local conductivity, $\Delta_U \sigma^i_{xx} = \sigma_{xx}^i(U=U_0) - \sigma_{xx}^i(U=0)$, are averaged for sites with similar local moments. 
    (a): Local conductivity [Eq.~\eqref{eq:sigma-i}] at $T$ vs.\ local moment measured. 
    (b): Interaction-induced corrections vs.\ magnetic moment. 
    The curves in (a) and (b) display a nearly $T$-independent correlation. 
    (c): The temperature $T_\sigma^*$ where the conductivity peaks,  
    the fractal-enhanced spin-flip scattering rate [Eq.~\eqref{eq:spin-flip}], 
    and the naive rate estimate $1/(\tau_s U_0) \sim \mathcal{F}_M$ 
    vs.\
    the average distance between dopant atoms. 
    The inset shows the Chalker exponent $\Upsilon=1-d_2/2$ with $d_2$ the second multifractal dimension. The data in (a) and (b) are obtained on a $100\times 100$ lattice and are averaged over $31$ disorder realizations, 
    and the data in (c) and (d) are averaged over $10$ disorder realizations.}
    \label{fig:sigma-siz-correlation}
\end{figure}

The correlation between transport and magnetism becomes best visible when grouping the lattice sites into bins 
of similar magnetization (e.g. $0.025\le |S_i^z|<0.125$).
We then average the local conductivity $\sigma^i_{xx}$ 
within each bin. Fig.~\ref{fig:sigma-siz-correlation}(a)[(b)]
shows the correlation between  
$\sigma^i_{xx}(T)$ 
[$\Delta_U \sigma^i_{xx}(T)$]
and sites with similar $|S_i^z|(T)$, 
which, interestingly, turns out to be nearly $T$-independent. 
As the sites progressively magnetize with decreasing temperature, 
the local conductivity first continues to increase a bit, but as soon as a moderate polarization ($|S_i^z(T=0)| \gtrsim 0.1$ is reached, the current flow is strongly affected and the average local conductivity diminishes significantly with further polarization.

Our HF calculations predict the nucleation of frozen moments already for $T \lesssim T_s^*$ $(> T_\sigma^*)$. 
In reality, these moments fluctuate~\footnote{However, the moments are quasi-static on the time-scale conduction electrons 
need to cross the few sites on which the moment is localized.}
until, at significantly lower temperatures, magnetic order (likely glassy) sets in. 
Nonetheless, the HF approximation correctly captures the essence of the MI crossover, as the key role of these moments is to induce spin-flip scattering of the current-carrying electrons. Assuming a 
Kondo-like coupling to local moments, we estimate the rate of quasi-elastic spin-flip carrier scattering 
to be of order (see End Matter)
\begin{align}\label{eq:spin-flip}
    {1}/{\tau_s}(T) \sim U_0 \left[ \mathcal{F}_M(T) \right]^{1/(1 + \chalk)}.     
\end{align}
Here, $\mathcal{F}_M$ is the fraction of sites occupied by moments, and 
$\chalk \equiv (2 - d_2)/2$, where $d_2$ is the second Renyi dimension characterizing 
the fractality of eigenstates within $1/\tau_s$ of the Fermi energy.
At $R= 2 a_0$ we find that $\chalk \approx 0.65$,
Fig.~\ref{fig:sigma-siz-correlation}(d). 
Strong spin-flip scattering eventually suppresses the triplet-interaction enhancement of conductance present 
at higher $T$, inducing an insulating trend that begins
at $T^*_\sigma$.
This is the temperature 
where 
$1 / \tau_s(T = T^*_\sigma) \sim T^*_\sigma$.
At colder temperatures, the triplet-channel AA corrections are neutralized 
by dynamical symmetry breaking. 
Fig.~\ref{fig:sigma-siz-correlation}(c) shows that $T^*_\sigma$
and $1/\tau_s(T^*_\sigma)$ [Eq.~\eqref{eq:spin-flip}] 
both increase monotonically and track each other for $R > 2 a_0$. 
Note that this occurs while the magnetic moments cover only a tiny fraction of sites. 
Were moments to occupy random positions, the scattering rate would be limited to 
$\sim U_0 \, \mathcal {F}_M \ll T_{\sigma}^*$. 
It is because both the magnetic moments and the wavefunction probabilities tightly cluster together
that the spin-flip scattering is strongly boosted relative to the naive expectation. 

The spatial correlations between moments and currents exhibited in Figs.~\ref{fig:sigmai} and \ref{fig:dsigmai}, quantified by Fig.~\ref{fig:sigma-siz-correlation}(a,b), along with the observation of $T_\sigma^* \sim 1 / \tau_s$ [Fig.~\ref{fig:sigma-siz-correlation}(c)], explains the physical mechanism of the MI crossover. Exchange-fueled AA enhancements channel stronger currents through a landscape otherwise dominated by weak localization  (since the bare $\sigma_{xx} \sim e^2/h$). At $T = T_s^*$, moments begin to form due to the runaway fractal enhancement of the triplet channel. 
At the threshold $1 / \tau_s(T = T_\sigma^*)\gtrsim T_{\sigma}^*$, spin-flip scattering overwhelms the current-stabilizing triplet channel, and the entire landscape succumbs to interaction-modulated Anderson localization.


\emph{Conclusion and Outlook}---In this Letter, we have used large-scale self-consistent HF calculations to demonstrate that the onset of magnetic moments induces a low-$T$ MI-crossover to insulating behavior   
in a model of a 2D $\delta$-layer. Our observations are consistent with the 40-year-old scaling theory \cite{Finkelstein1983,BK1994} in that the conductivity and density of states are initially enhanced as temperature is lowered; we attribute this to the disorder-amplified spin-triplet interaction channel. 
By calculating and correlating local quantities, we have shown that a sparse set of moments form preferentially on the fractal support where the strongest currents flow. A negative feedback effect thereby ensues, wherein the concomitant breaking of spin SU(2) symmetry 
disintegrates  
the spin-triplet channel and suppresses the most conducting paths. 

Our work opens up many avenues for future exploration. 
A key question 
concerns the presence and nature of genuine magnetic order as $T \to 0$ \cite{Zhang2024}, and its effect on magnetotransport~\cite{danna2025}. The degree of frustration among effective couplings between local clusters
will decide whether a spin glass emerges (predicted for strong enough frustration~\cite{viteritti2026}) or a competing phase, such as a random valence bond state~\cite{Bhatt1982}.

Another question is whether the local correlations between currents and magnetic fluctuations can be extracted from  Finkel'stein's approach, 
considering higher-moment correlators of both currents and spins 
\cite{Lerner1988,Altshuler1991,Burmistrov2013,Richardella2010}.
Finally, it would be edifying to understand the similarities and differences between the MI crossover observed here and in gate-tuned ones  in silicon MOSFETs, occurring at much lower densities
\cite{Kravchenko1994, Kravchenko1995, Abrahams2001, Kravchenko2010, Shashkin2017}. While local magnetism and spin glass-like correlations probably play similar roles, MOSFETs may well display additional interesting features due to the more prominent role of long-range Coulomb interactions and the ensuing frustration in the charge sector~\cite{Ovadyahu2013,Amini2014}.

\acknowledgements
Part of this work was performed at the Aspen Center for Physics, which is supported by National Science Foundation Grant No. PHY-2210452. The project was supported by 
the ERC Horizon 2020 research and innovation program, Grant Agreement 810451 (HERO). 
X.Z. acknowledges support from Paul Scherrer Institute, and from the
ERC under the EU’s Horizon 2020 research and innovation program, Grant Agreement 810451. 
X.Z. also acknowledges the Olle Eriksson foundation and Roland Gustafssons Stiftelse för teoretisk fysik for travel support. This work was supported in part by the NOTS cluster operated by Rice University's Center for Research Computing (CRC). Part of the computational resources were provided by the National Academic Infrastructure for Supercomputing in Sweden (NAISS), funded by the Swedish Research Council.

\bibliography{ref}

\appendix

$\phantom{0}$\\
\centerline{{\bf End Matter}}
\vspace{2pt}

\emph{Appendix A: Finkel'stein RG and magnetic instability}---Virtual renormalization effects in the weakly disordered, two-dimensional diffusive Fermi liquid are encoded via the beta functions \cite{Finkelstein1983,Castellani1984,BK1994}
\bsub\label{FRG}
\begin{align}
    \frac{d \sigma}{d l}
    =&\,
    -
    \frac{2}{\pi}
    -
    \frac{2}{\pi}
    \left[
        1 - \left(1 - \frac{1}{\gamma_s}\right)\ln(1 - \gamma_s)
    \right]
\nonumber\\
&\,
   -
    \frac{6}{\pi}
    \left[
        1 - \left(1 - \frac{1}{\gamma_t}\right)\ln(1 - \gamma_t)
    \right],
\label{betasigma}
\end{align}
\begin{align}
    \frac{d \gamma_s}{d l}
    =&\,
    -
    \frac{2}{\pi \sigma}
    (1 - \gamma_s)(\gamma_s + 3 \gamma_t),
\label{betas}
\\
    \frac{d \gamma_t}{d l}
    =&\,
    \frac{2}{\pi \sigma}
    (1 - \gamma_t)(\gamma_t - \gamma_s),
\label{betat}
\\
    \frac{d \ln \nu}{d l}
    =&\,
    \frac{2}{\pi \sigma}
    \left[
        \ln(1 - \gamma_s) + 3 \ln(1 - \gamma_t)
    \right].
\label{betados}
\end{align}
\esub
Here $l$ denotes the logarithm of the RG length scale $L$,
$\sigma = \sigma_{dc} \, h / e^2$ is the average dimensionless conductivity, 
$\gamma_{s}$ and $\gamma_t$ respectively denote the 
dimensionless density-density (spin singlet) and spin-spin exchange (triplet) interaction strengths, 
while
$\nu$ is the density of states (DoS) per spin evaluated at the Fermi level. 

The parameters 
$
    \gamma_{s,t} 
    =
    {F_{s,t}}/{(1 + F_{s,t})},
$
where $F_{s,t} = 2 \, \nu_0 \, U_{s,t}$ is the singlet (triplet) Fermi liquid parameter. 
These coupling strengths should be understood as \emph{spatial averages}; all observables, including
quantum expectation values of local operators such as four-fermion interactions,
acquire nontrivial probability distributions in a quantum system with quenched disorder. 
In the absence of Anderson localization, distributions can be characterized by disorder-averaged moments \cite{Evers2008};
distributions typically become too broad to possess higher moments in an Anderson insulator \cite{AltshulerPrigodin1989}.
The exchange coupling $U_t \sim - e^2/k_F$ is negative, so that $F_t < 0$; $F_t = -1$ 
($\gamma_t \rightarrow -\infty$) represents the Stoner instability to itinerate ferromagnetism. 
In Eq.~(\ref{FRG}), we ignore the BCS pairing instability. 

The first, second, and third terms in Eq.~(\ref{betasigma})
represent the weak localization, singlet- and triplet-channel
Altshuler-Aronov (AA) corrections \cite{AA1985,Lee1985}. Since $\gamma_t < 0$, 
the triplet correction is \emph{anti-localizing} \cite{Finkelstein1983,BK1994}, 
opposing the localizing effect in the singlet channel ($\gamma_s>0$); 
related AA corrections to the DoS appear in Eq.~(\ref{betados}). 
The interaction renormalizations in Eqs.~(\ref{betas}) and (\ref{betat}) 
arise in a diffusive Fermi liquid due to wave-function multifractality \cite{Evers2008,Feigelman2007,Foster2014},
i.e.\ the critical scaling of \emph{moments of the local DoS} 
due to the disorder.

Under the RG, $\gamma_s$ flows to 1 (the incompressible limit in the density-density channel \cite{BK1994}),
while the triplet coupling diverges, $\gamma_t \rightarrow - \infty$, at a finite renormalization scale. 
The divergence translates into a finite enhancement of the conductivity, 
\begin{align}
    \ln\left[
        \frac{
                \sigma(\gamma_t \rightarrow - \infty)
            }{
                \sigma(\gamma_t^\pup{0})
            }
    \right]
    \simeq
    3 
    \int_{|\gamma_t^\pup{0}|}^\infty d y \, \frac{\ln(y)}{y^2},
\end{align}
so that $\sigma^* \equiv \sigma(\gamma_t \rightarrow - \infty)$ 
is non-critical, neither infinite nor zero. 
Taking the dynamic critical exponent for temperature scaling $z_T \sim 2$
(which holds until very close to the instability), 
the RG predicts a magnetic instability at the temperature 
\begin{align}
    T^*_s
    \simeq
    T_0
    \,
    \exp\left[
        -
        \frac{\pi \sigma^* (1 - |F_t^\pup{0}|)}{|F_t^\pup{0}|}
    \right],
\end{align}
where $|F_t^\pup{0}| < 1$ is the triplet Fermi-liquid parameter
at the initial temperature scale $T_0 \sim 1/\tau_{\rm el} > T$
($\tau_{\rm el}$ is the elastic impurity scattering time).

\begin{figure}[t]
    \centering
    \includegraphics[width=0.95\linewidth]{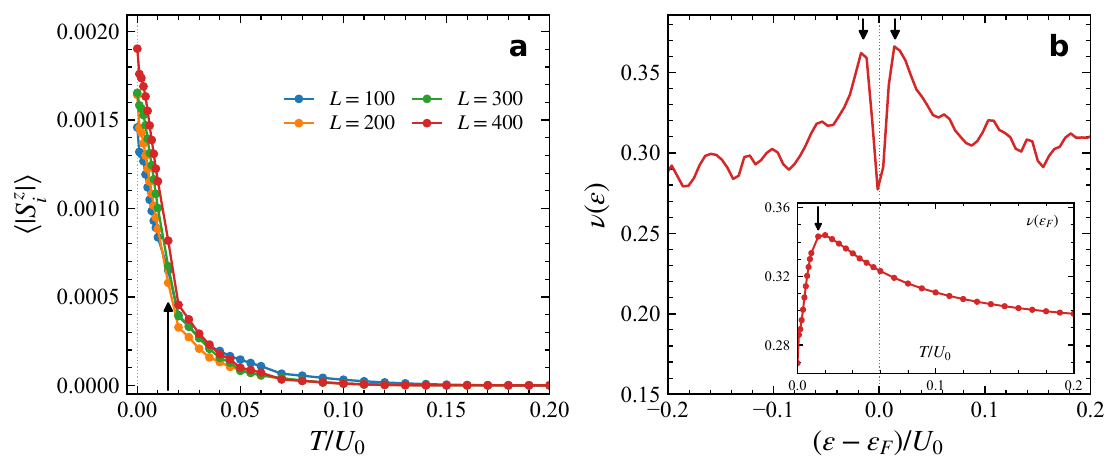}
    \caption{(a) The local moment density 
    versus temperature $T$ for different system sizes $L$. The  arrow indicates $T_\sigma^*$ where $\sigma_{xx}(T)$ peaks.
    (b) The DoS $\nu(\varepsilon)$ versus energy for a system of linear size $L=400$ having the nearly critical density $R=2a_0$, at $T = 0$.  The black arrows indicate the excitation energy $\varepsilon-\varepsilon_F=T_\sigma^*$.
    The inset shows the $T$-dependence of $\nu(\varepsilon_F)$, displaying the AA enhancement above $T_\sigma^*$ (indicated by the arrow) and the corresponding suppression below.
    }
    \label{fig:Lscaling}
\end{figure}

Although $T^*_s$ would seem to indicate a Stoner transition to ferromagnetism, 
our numerics suggests a more subtle picture. Due to spatial inhomogeneity, relatively rare regions most prone to a Stoner instability start forming local moments. While those 
still fluctuate on long time scales, they induce an effective interaction-mediated reduction of the spin SU(2) symmetry on the timescale relevant for the itinerant electrons; this happens precisely in the regions where the DoS and DC conductivity are strongest. Further regions magnetize only at lower temperature, wherein we found the growth of magnetic moments to correlate with a significant reduction of the local conductivity, cf.\ Fig.~\ref{fig:sigma-siz-correlation}. At yet lower temperature the mutual (possibly RKKY-like) coupling between magnetic clusters becomes relevant.
Small easy-axis or easy-plane spin anisotropy can then stabilize
long-range or quasi-long-range magnetic order or --quite likely-- spin glass-type freezing. 

The effective reduction of spin SU(2) to U(1) symmetry by moment formation 
dramatically alters the 
trajectory of the subsequent RG flow upon further lowering the temperature. 
In particular, the antilocalizing AA correction on the second
line of Eq.~(\ref{betasigma}) has its prefactor reduced from $6/\pi$ to 
$2/\pi$; the combination of weak localization and the singlet-channel AA 
on the first line is 
expected to dominate
and the flow is towards an interacting Anderson insulator at zero temperature. 
The latter is assured if the combination of broken spin and time-reversal symmetries persists down to $T=0$. 
However, we cannot fully exclude the scenario (beyond our HF treatment) in which magnetic moments get Kondo screened 
at very low temperatures, reinstalling both symmetries (for a review, see e.g.~\cite{BK1994}).
Such a scenario would seem hard to reconcile with magnetotransport and hysteresis effects 
reported in recent experiments, at least for low enough densities~\cite{danna2025}.

As $\gamma_t \rightarrow - \infty$, the DoS at the Fermi level
is generically enhanced. For sufficiently large $|\gamma_t^\pup{0}| \gtrsim 1$, 
the asymptotic behavior is
\begin{align}
    \nu(T) 
    \simeq&\,
    \nu(T_0)
    \,
    \left(\frac{T_0}{T}\right)^{\eta},\quad T> T^*_{s},
\nonumber\\
    \eta
    \simeq&\,
    \frac{1}{\pi \sigma^*}
    \left\{
        \ln(1 - \gamma_s^\pup{0})
        +
        3
        \ln\left[|\gamma_t^\pup{0}|\right]
    \right\}
    > 0.
    \label{eq:dos}
\end{align}
Below the temperature $T^*_{\sigma}$ {where the magnetic moments start dominating the IR cut-off of the AA corrections}, the scaling of the DoS  is expected to eventually be dominated by the singlet channel, and result in a power $ \nu(T) \sim T^{\eta'}$ with positive $\eta'>0$. 
Accordingly, the $T=0$ DoS is expected to exhibit a pseudogap $\nu(\epsilon)\sim \epsilon^{\eta'}$, raising up to a peak around $\epsilon \equiv E-E_F \sim T^*_\sigma$, and then decreasing again at higher energy. 
{This is consistent with our numerical results obtained for the DoS, shown in the main panel of Fig.~\ref{fig:Lscaling}(b).
The DoS at the Fermi level is plotted versus temperature in the inset to Fig.~\ref{fig:Lscaling}(b). It exhibits a peak near the temperature $T^*_{\sigma}$, as expected from the RG flow discussed above.
Above $T^*_{\sigma}$, the spin-triplet channel interaction dominates and produces a positive correction to the DoS [Eq.~\eqref{eq:dos}]. Below $T^*_{\sigma}$, the Fermi-level DoS is suppressed, suggesting that the spin-singlet channel interaction dominates.
}

\emph{Appendix B: Spin-flip scattering rate}---We estimate the rate of spin-flip scattering induced by a dilute
set of magnetic moments on the (near-)critical itinerant electrons. At $R = 2 a_0$, 
the interaction and hopping energy scales are both of order $U_0$. Then the rate of spin-flip scattering
must scale proportional to $U_0$, 
\begin{align}\label{tau_s_Def}
    \frac{1}{\tau_s}
    \sim
    U_0
    \,
    \mathcal{F}_M',
\end{align}
where $\mathcal{F}_M'$ is the typical probability of a carrier to encounter a magnetic moment at the next site it visits, and to scatter from it.

The moments and critical wave functions both strongly concentrate where the triplet interaction 
has been most enhanced. This amplifies the scattering probability 
\begin{align}\label{FMprimeDef}
    \mathcal{F}_M' \sim \mathcal{F}_M \left({L^*}/{a}\right)^{2 \Upsilon},
\end{align}
where $2 \Upsilon = 2 - d_2$ is (twice) the Chalker scaling exponent \cite{Chalker1990,Feigelman2007}, 
$a$ is the lattice spacing, and $L^*$ is a length scale discussed below. 
The magnetic instability that forms the moments also arises from the same 
``pre-localization'' mechanism 
(multifractality, encoded via the Renyi dimension $d_2 > 0$),
leading to a relevant scaling dimension $2 \Upsilon$ for the triplet interaction \cite{Foster2014}.
This should be contrasted with the case of a clean Fermi liquid, where triplet interactions
are purely marginal \cite{Shankar1994} and a Stoner instability occurs only for sufficiently
strong bare magnetic exchange. 

The length scale $L^*$ in the spin-flip scattering rate should be self-consistently limited,
since carriers survive a duration $\tau_s$ before flipping:
\begin{align}\label{LstarDef}
    \frac{L^*}{a}
    \sim
    \sqrt{\frac{D \, \tau_s}{a^2}}
    \sim
    \sqrt{U_0 \, \tau_s}.
\end{align}
Here we use the Einstein relation $D=\sigma/ 2 \pi \nu_0$. At the MI crossover $\sigma \sim 1$,
and for $R = 2 a_0$ the DoS $\nu_0 \sim 1 / U_0 \, a^2$. 
Combining Eqs.~(\ref{tau_s_Def})--(\ref{LstarDef}) gives Eq.~(\ref{eq:spin-flip}).

\end{document}


\title{Supplemental Material: 
Fractal Confinement and Magnetic Self-sabotage of Current Flow: Electrons Near the Metal-insulator Crossover in a 2D $\delta$-layer}
\author{Xinghai Zhang}
\affiliation{Department of Physics, KTH Royal Institute of Technology, SE-106 91 Stockholm, Sweden}
\author{Matthew S. Foster}
\affiliation{Department of Physics and Astronomy, Rice University, Houston, Texas 77005, USA}
\author{Markus M\"uller}
\affiliation{PSI Center for Scientific Computing, Theory and Data, CH-5232 Villigen PSI, Switzerland}

\date{\today}

\maketitle
\tableofcontents

\section{Details on the model of a $\delta$-layer}\label{sec:model}

\subsection{Couplings of the Hubbard model}

For completeness we rederive below the expressions for hopping integrals and onsite repulsion for hydrogenic dopant wavefunctions, cf. e.g. Ref. \cite{MottDavis1979}
.  
The Hamiltonian of an electron hopping between two dopant atoms, one located at the origin and one at position $\mathbf{R}$, is given by
\begin{equation}
    H =
    -\frac{\nabla^{2}}{2m}
    -\frac{e^{2}}{\kappa r}
    -\frac{e^{2}}{\kappa \left|\mathbf{r}-\mathbf{R}\right|}\,.
    \label{eq:H_2atom}
\end{equation}
Here $m$ is the effective mass of the electron and $\kappa$ is the dielectric constant of the host semiconductor.
The tight-binding hopping amplitude is then given by
\begin{equation}
    t= - \intop_{\mathbf{r}}\phi^{*}\left(\mathbf{r}\right)H\phi\left(\mathbf{r}-\mathbf{R}\right)\,,
    \label{eq:hopping}
\end{equation}
where $\phi(\mathbf{r})$ is the ground state wavefunction of the electron on a single dopant. 
Here we consider the hydrogenic approximation for the atomic orbital,
\begin{equation}
    \phi(\mathbf{r}) = \frac{1}{\sqrt{\pi}a_0^{3/2}} e^{-r/a_0}\,,
\end{equation}
where $a_0$ is the Bohr radius of the hydrogen-like atom. 
Thus, we have 
\begin{equation}
    t = 
    - E_{0}\intop_{\mathbf{r}}\phi\left(\mathbf{r}-\mathbf{R}\right)\phi\left(\mathbf{r}\right)
    +
    \intop_{\mathbf{r}}\phi\left(\mathbf{r}-\mathbf{R}\right)\frac{e^{2}}{\kappa\left|\mathbf{r}-\mathbf{R}\right|}\phi\left(\mathbf{r}\right),
\end{equation}
where $E_0 = -e^2/2\kappa a_0$ is the binding energy of the valence electron.
Evaluating the integrals,
\begin{align}
    \intop_{\mathbf{r}}\phi\left(\mathbf{r}-\mathbf{R}\right)\phi\left(\mathbf{r}\right)
    &=
    \frac{1}{\pi a_0^3}\intop_{\mathbf{r}}e^{-r/a_0-\left|\mathbf{r}-\mathbf{R}/a_0\right|} = \left(1+\frac{R}{a_0} + \frac{R^2}{3a_0^2}\right)e^{-R/a_0}\,,\\
    \intop_{\mathbf{r}}\phi\left(\mathbf{r}-\mathbf{R}\right)\frac{1}{\left|\mathbf{r}-\mathbf{R}\right|}\phi\left(\mathbf{r}\right)
    &= \frac{1}{a_0} \left(1 + \frac{R}{a_0}\right) e^{-R/a_0}\,,
\end{align}
we find for the hopping between two dopants
\begin{align}
\label{eq:tRij}
    t &= t_0 e^{-R/a_0}\,, \quad
    t_0 = \frac{e^2}{\kappa a_0} \left[\frac{3}{2}\left(1+\frac{R}{a_0}\right) + \frac{R^2}{6a_0^2} \right]\,.
\end{align}

The onsite Hubbard repulsion between two electrons residing on the same dopant ion is given by
\begin{equation}
    U_0
    =\intop_{\mathbf{r}_{1}\mathbf{r}_{2}}\frac{e^{2}}{\kappa r_{12}}\left|\phi\left(\mathbf{r}_{1}\right)\right|^{2}\left|\phi\left(\mathbf{r}_{2}\right)\right|^{2}\,.
\end{equation}
For hydrogen-like wavefunctions this yields
\begin{align}
    U_0&=\frac{e^{2}}{\kappa}\frac{1}{\pi^{2}a_0^6}\intop_{\mathbf{r}_{1}\mathbf{r}_{2}}e^{-2(r_{1}+r_{2})/a_0}\frac{1}{\sqrt{r_{1}^{2}+r_{2}^{2}-2r_{1}r_{2}\cos\theta_{12}}} 
    = \frac{5e^2}{8\kappa a_0}\,.
\end{align}

\begin{figure}
    \centering
    \includegraphics[width=0.8\linewidth]{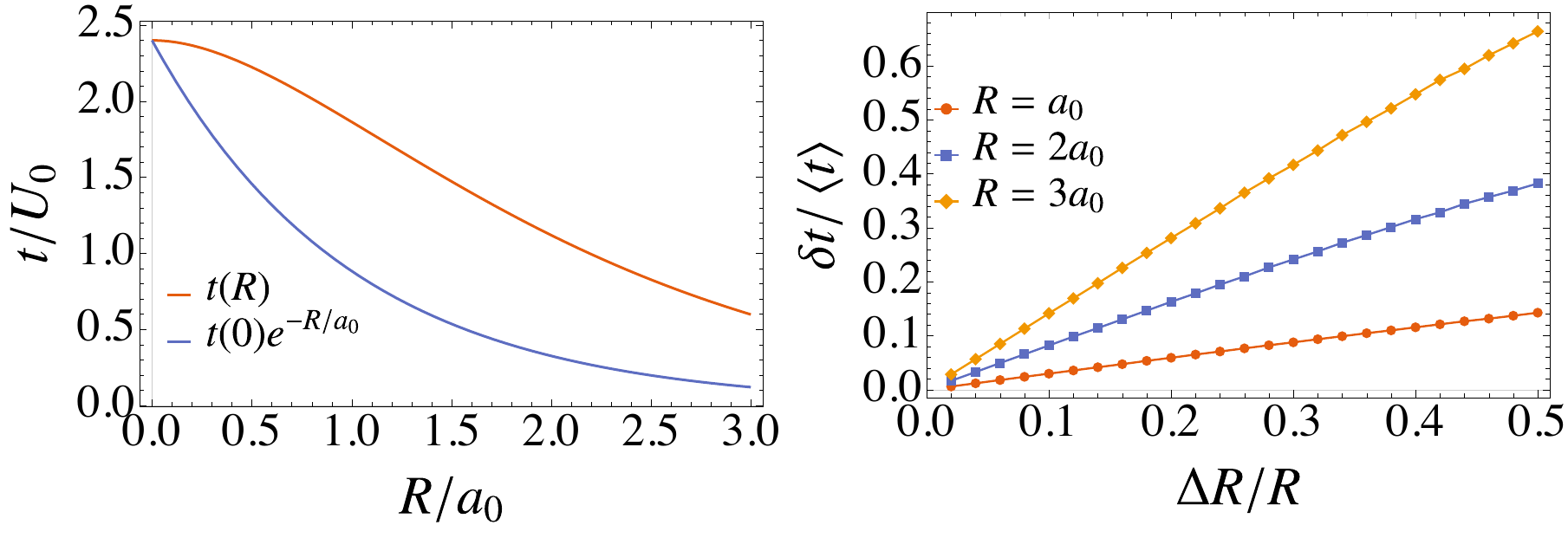}
    \caption{
    The left panel shows the 
    $R$-dependence of the hopping amplitude. 
    Note that $t(R)$ (red) decays with $R$ more slowly than a pure exponential (blue curve) due to the prefactor in Eq.~(\ref{eq:tRij}). 
    The right panel depicts
    the disorder-induced spread of amplitudes 
    $\delta t = \sqrt{\langle t_{ij}^2 \rangle - \langle t_{ij} \rangle^2}$
    versus $\Delta R / R$. Here $\Delta R$ is the radius of a disk centered at $R$.
    Each site is randomly displaced according to a uniform distribution over the disk. 
    }
    \label{fig:t-dt-R}
\end{figure}

\subsection{Simulating a positionally random $\delta$-layer}

In our simulations, we employ a tight-binding model of dopant atoms that are nearly Poissonian distributed in the 2D plane. To generate the random lattice, we start from a triangular lattice with spacing $R$ and randomly displace every atom, sampling a position uniformly within a disk of radius $\Delta R$ centered at its regular lattice position. For the tight-binding terms we only retain the $6$ nearest neighbor bonds of the underlying triangular lattice, but determine the individual hopping amplitudes from the actual distance $R_{ij}$ between the corresponding lattice sites, as specified in Eq.~\eqref{eq:tRij}. The $R$-dependence of the hopping amplitude $t$ and the relative variation of the hoppings, $\delta t/t$, due to the positional disorder are shown in Fig.~\ref{fig:t-dt-R}. 
The variations in the hopping amplitude $\delta t$ increase approximately linearly with the relative displacement amplitude $\Delta R/R$, which quantifies the deviation from the regular triangular lattice. For $\Delta R= 0.5 \, R$,
the distorted triangular lattice constitutes a good approximation for Poisson-distributed dopants, and we stick to this choice for the main part of the paper.
Note that the relative variation of the hoppings increases not only with $\Delta R/R$, but also with  the average dopant spacing $R$. Therefore, as the doping density is lowered, the system becomes both more disordered and more correlated, as measured by the larger relative interaction strength $U_0/\langle t\rangle$.


\section{Linear response theory of local transport}\label{sec:transport}

\subsection{Kubo formula for the local conductivity: Local current density response to a uniform electric field}

We employ textbook linear response theory to evaluate the local current response to a uniform electric field. {Within linear response we may treat the Hartree-Fock Hamiltonian as an effective Hamiltonian of non-interacting electrons. Below, we briefly rederive the linear response of non-interacting electrons in homogeneous space, which then carries over with little  modification to lattices with onsite potential and bond-dependent hoppings.} 

In the presence of a time-dependent vector potential the Hamiltonian of free fermions reads,
\begin{equation}
H=\intop_{\mathbf{x}}\psi^{\dagger}
\left[\frac{\left(-i\nabla-e\mathbf{A}\right)^{2}}{2m} {- \mu}
\right]
\psi\,,
\end{equation}
where $\mu$ is the chemical potential ensuring half filling, and we use the unit with $\hbar=1$.

The number current density is then given by,
\begin{align}
\mathbf{j}(\mathbf{x}) & =-\frac{\delta H}{\delta e \mathbf{A}\left(\mathbf{x}\right)} = \mathbf{j}^{P}(\mathbf{x})+\mathbf{j}^{D}(\mathbf{x})\,,\\
\mathbf{j}^{P}(\mathbf{x}) & = - \frac{i
}{m}\psi^{\dagger}\left(\mathbf{x}\right)\overleftrightarrow{\nabla}\psi\left(\mathbf{x}\right)\,,\qquad
\mathbf{j}^{D}(\mathbf{x})= - \frac{e
\mathbf{A}}{m}\psi^{\dagger}\left(\mathbf{x}\right)\psi\left(\mathbf{x}\right)\nonumber\,.
\end{align}
Here $\overleftrightarrow{\nabla}=\left(\overrightarrow{\nabla}-\overleftarrow{\nabla}\right)/2$. $\mathbf{j}^P$, $\mathbf{j}^D$ and $\mathbf{j}$ are the paramagnetic, diamagnetic and total 
number current density, respectively.
We work in the London gauge and the electric field is encoded in the time-dependent vector potential $\mathbf{A}(t)$ via $\mathbf{E} = -\partial_t \mathbf{A}$. The vector potential is assumed to be turned on slowly, with $\lim_{t\to -\infty} \mathbf{A}(t) = 0$.

Using standard 
linear response theory, the local current density response to a uniform external field is given by, 
\begin{align}
    j_{\alpha}\left(\mathbf{x},t\right) & =-\sum_\beta \left[\int_{t'}
    \Pi^R_{\alpha\beta}
    \left(\mathbf{x};t,t'\right)
    {e}A_{\beta}\left(t'\right)
    +\frac{n\left(\mathbf{x}\right) 
    }{m}\delta_{\alpha\beta}\,
    {e}A_{\beta}\left(t\right)\right]\,,\\
    \Pi^R_{\alpha\beta}
    \left(\mathbf{x};t,t'\right) 
    & =-i \left\langle\left[j_{\alpha}^{P}\left(\mathbf{x},t\right),\mathcal{J}_{\beta}^{P}\left(t'\right)\right]\right\rangle
    \Theta\left(t-t'\right)\,.
\end{align}
{Here $\mathcal{J}^P_\beta = \int_\mathbf{x} j^P_\beta(x)$ is the volume integral of the} {number current density, which couples linearly to a spatially homogeneous vector potential $e{\bf A}$;  $n\left(\mathbf{x}\right)= \psi^{\dagger}\left(\mathbf{x}\right)\psi\left(\mathbf{x}\right)$ is the number density.}
In the London gauge, we have $\mathbf{E}=-\partial_{t}\mathbf{A}$, or in frequency space, 
\begin{equation}
\mathbf{A}(\Omega,\mathbf{x})=\frac{\mathbf{E}(\Omega,\mathbf{x})}{i\Omega}\,.
\end{equation}
Turning to frequency space, the local conductivity, defined by ${e}j_\alpha\left(\Omega, \mathbf{x}\right)= \sum_\beta \sigma_{\alpha\beta}\left(\Omega, \mathbf{x}\right) \mathbf{E}_\beta(\Omega,\mathbf{x})$, is  given by
\begin{equation}
    \sigma_{\alpha\beta}\left(\Omega, \mathbf{x}\right)
    =
    - 
\frac{e^2}{i\Omega}\left[\Pi_{\alpha\beta}\left(\Omega, \mathbf{x}\right) + \frac{n\left(\mathbf{x}\right)
}{m}\delta_{\alpha\beta}\right]\,.
\end{equation}
In the imaginary-time formulation, the current-current correlation function
can be written as 
\begin{equation}
    \Pi_{\alpha\beta}
    \left(i\Omega_{m}, \mathbf{x}\right)
    =T\sum_{\omega_{n}}\mathsf{Tr}\left[j^P_{\alpha}(\mathbf{x})G\left(i\omega_{n}\right)\mathcal{J}^P_{\beta}G\left(i\omega_{n}+i\Omega_{m}\right)\right]\,.
\end{equation}
Here, $T=\beta^{-1}$ is the temperature, and $\omega_{n}=\left(2n+1\right)\pi/\beta$ and $\Omega_{m}=2m\pi/\beta$
are the fermionic and bosonic Matsubara frequencies, respectively. The
Green's function is given by,
\begin{equation}
G\left(i\omega_{n}\right)=\frac{1}{-i\omega_{n}+H}\,.
\end{equation}
Evaluating the Matsubara summation, we have
\begin{equation}
    \Pi_{\alpha\beta}(i\Omega_m, \mathbf{x}) 
    =
    \int_{-\infty}^{\infty}d\varepsilon
    f(\varepsilon)
    \mathsf{Tr}\left\{
    j^P_{\alpha}(\mathbf{x})
    G\left(\varepsilon+i\Omega_{m}\right)
    \mathcal{J}^P_{\beta}
    \delta(\varepsilon - H)
    +
    j^P_{\alpha}(\mathbf{x}) 
     \delta(\varepsilon - H)
    \mathcal{J}^P_{\beta}
    G\left(\varepsilon-i\Omega_{m}\right)
   \right\} \,,
\end{equation}
with the Fermi-Dirac distribution 
$f\left(\varepsilon\right)= (e^{\beta \varepsilon}+1)^{-1}$. 
The retarded current-current correlation function can be obtained
by analytical continuation,
\begin{align}
    \Pi_{\alpha\beta}^{R}\left(\Omega\right) 
    & =-\Pi_{\alpha\beta}\left(i\Omega_{m}\to\Omega+i\eta\right)\nonumber \\
    & =\int_{-\infty}^{\infty}d\varepsilon   f(\varepsilon)\mathsf{Tr}\left\{ j^P_{\alpha}(\mathbf{x})
    G_{R}\left(\varepsilon+\Omega\right)
    \mathcal{J}^P_{\beta}
    \delta(\varepsilon - H)
    +
    j^P_{\alpha}
    \delta(\varepsilon - H)
    \mathcal{J}^P_{\beta}
    (\mathbf{x})G_{A}\left(\varepsilon-\Omega\right)
    \right\} \,.
\end{align}
Here the retarded/advanced Green's function is given by,
\begin{equation}
G_{R/A}\left(\omega\right)=-G\left(i\omega_{n}\to\omega\pm i\eta\right)\,.
\end{equation}
Due to gauge invariance, the DC conductivity should be a finite quantity, the $\Pi_{\alpha\beta}\left(\Omega=0\right)$
term must exactly cancel with the diamagnetic term. Thus we have the local conductivity, 
\begin{equation}
    \sigma_{\alpha\beta}
    \left(\Omega, \mathbf{x}\right)
    =
    -\frac{e^2}{i\Omega}
    \left[
    \Pi_{\alpha\beta}^{R}\left(\Omega, \mathbf{x}\right)
    -
    \Pi_{\alpha\beta}^{R}\left(0, \mathbf{x}\right)
    \right]\,.
\end{equation}
We then reach the Kubo-Bastin-type formula for local DC conductivity, 
\begin{equation}
    \sigma_{\alpha\beta}^{\mathsf{DC}}(\mathbf{x})
    =\sigma_{\alpha\beta}\left(\Omega\to0, \mathbf{x}\right)
    =i{e^2}{}\int_{-\infty}^{\infty}d\varepsilon f\left(\varepsilon\right)\mathsf{Tr}
    \left[j^P_{\alpha}(\mathbf{x})
    G_{R}'\left(\varepsilon\right)
    \mathcal{J}^P_{\beta}
    \delta\left(\varepsilon-H\right)    
    -
    j^P_{\alpha}(\mathbf{x})
    \delta\left(\varepsilon-H\right)
    \mathcal{J}^P_{\beta}
    G_{A}'\left(\varepsilon\right)
    \right] \,.
\end{equation}
Below, we use the notation $\sigma_{\alpha\beta}$ for the DC conductivity. 

\subsection{Global, average Conductivity}
The global conductivity describes the {\em average} current density response to an external field. It is obtained by averaging the local conductivity,
\begin{equation}
    \sigma_{\alpha\beta} = \frac{1}{\mathcal{A}} \intop_\mathbf{x} \sigma_{\alpha\beta}(\mathbf{x}).
\end{equation}
Here $\mathcal{A}$ is the volume of the system.
In this way we recover the usual Kubo-Bastin formula for the conductivity, 
\begin{equation}
    \sigma_{\alpha\beta}\left(\mu,T\right)
    =i\frac{e^2}{\mathcal{A}}\int_{-\infty}^{\infty}d\varepsilon f\left(\varepsilon\right)\mathsf{Tr}
    \left[\mathcal{J}^P_{\alpha}
    G_{R}'\left(\varepsilon\right)
    \mathcal{J}^P_{\beta}
    \delta\left(\varepsilon-H\right)
    -
    \mathcal{J}^P_{\alpha}
    \delta\left(\varepsilon-H\right)
    \mathcal{J}^P_{\beta}
    G_{A}'\left(\varepsilon\right)
    \right] \,.
\end{equation}
The longitudinal conductivity can be further simplified to the Kubo-Greenwood formula, 
\begin{equation}
    \sigma_{xx}\left(\mu,T\right)
    =    \int d\,(\delta\mu)\left(
    -\frac{\partial f}{\partial\, \delta\mu}\right)
    \sigma_{xx}\left(\mu+\delta\mu,T=0\right)\,,
    \label{eq:sigmaT}
\end{equation}
where $\sigma_{xx}\left(\mu+\delta\mu,T=0\right)$ is the d.c. conductivity at  $T=0$, evaluated for the same effective Hamiltonian, but with chemical potential shifted to $\mu+\delta\mu$,
\begin{equation}
    \sigma_{xx}\left(\mu+\delta\mu,T=0\right)
    =
    \frac{\pi e^{2}}{\mathcal{A}}
    \mathsf{Tr}\left[\mathcal{J}^P_{x}\delta\left(\delta\mu-H\right)\mathcal{J}^P_{x}\delta\left(\delta\mu -H\right)\right]\,.
    \label{eq:sigmaE}
\end{equation}


\subsection{Generalization to a  lattice}

On a lattice, we associate a current density to the sites $i$ and the Wigner-Seitz cell surrounding them. The (smeared out) {paramagnetic} current density around site $i$ sums all the currents that flow on bonds emanating from that site, and distributes it over the Wigner-Seitz cell, suggesting the definition
\begin{equation}
    \mathbf{j}^P_i
    = 
    \frac{1}{2\mathcal{A}_i}
    \sum_j 
    \left[
    -i(\mathbf{r}_i - \mathbf{r}_j) t_{ij} c^\dagger_i c_j  +{\rm H.c.}
    \right]\,.
\end{equation}
Here $\mathcal{A}_i$ is the volume 
of the Wigner-Seitz cell containing the site $i$. The integrals over $\mathbf{r}$ in the Kubo formulae derived above are then replaced by site summations with the measure $\mathcal{A}_i$
\begin{equation}
    \intop_{\mathbf{r}} \to \sum_i \mathcal{A}_i
\end{equation}
The volume integral of the current density then becomes 
\begin{equation}
    \mathcal{J}^P_\alpha = \sum_{i<j} \left[
    -i(\mathbf{r}_i - \mathbf{r}_j)_\alpha t_{ij} c^\dagger_i c_j  +{\rm H.c.}
    \right] \,.
    \label{eq:J_tot}
\end{equation}
With the definition of $\mathbf{j}^P_i$, we obtain the local conductivity, describing the local current response, as given in Eq.~(4) in the main text.

\subsection{Bond current response to an external field}

Similarly, we may consider the current flowing on the bond $(ij)$ in response to a uniform external field.
The {number} current operator on the bond $(ij)$ is naturally defined by
\begin{equation}
I^P_{ij}=-i {t_{ij}}{}c_{i}^{\dagger}c_{j}+{\rm H.c.}\,. \label{eq:Iij}
\end{equation}

The electrical current  in linear response to a uniform external field is given by
\begin{equation}
    {e}\left< I_{ij} \right>
    =\sum_{\alpha=x,y}\Lambda_{ij;\alpha}E_{\alpha}\,,
\end{equation}
with the response function 
\begin{equation}
    \Lambda_{ij;\alpha}
     = i e^2 \int d\varepsilon\, f(\varepsilon)\, \mathsf{Tr}\!\left[
        I^P_{ij} G_R'(\varepsilon) \mathcal{J}^P_\alpha \delta(\varepsilon-H) 
        -
        I^P_{ij} \delta(\varepsilon-H) \mathcal{J}^P_\alpha G_A'(\varepsilon)
    \right].
    \label{eq:Lambda_ij}
\end{equation}

\section{Numerical Method}\label{sec:method}

\begin{figure}
    \centering
    \includegraphics[width=0.95\linewidth]{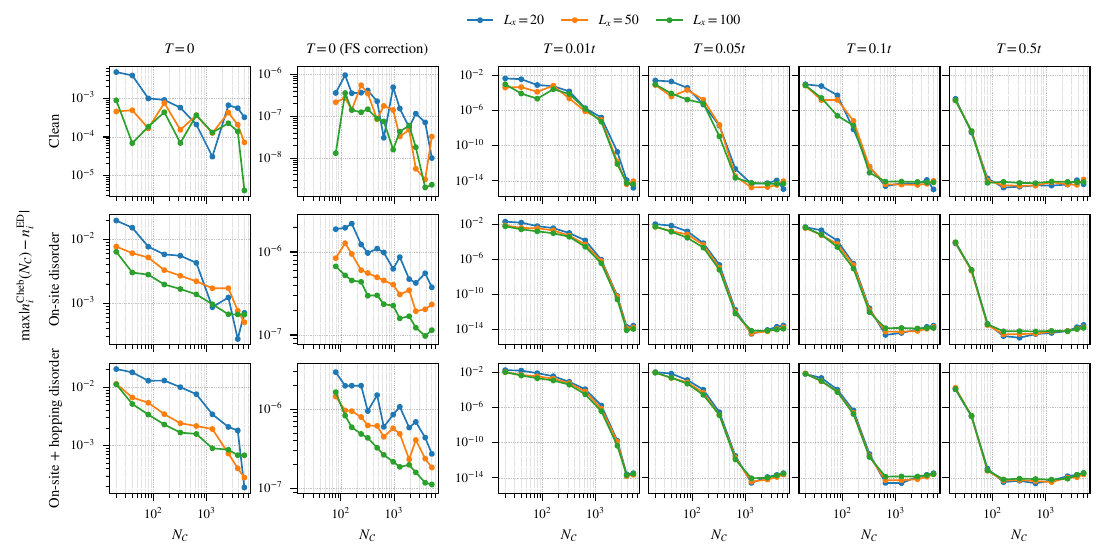}
    \caption{The maximal error in evaluating $n_i$ by the Chebyshev expansion 
    without using damping kernels, extracted from the comparison with exact diagonalization (ED). Results are shown as a function of the truncation $N_{\cal C}$. At finite $T$ exceeding the single particle level spacing, the error is nearly  independent of system size (right side of the panel). The truncation needs to be chosen roughly proportional to $N_{\cal C}\propto t/T$ to resolve the smeared Fermi edge and achieve a constant accuracy with lowering $T$. For strictly $T=0$ (left side of the panel), the states closest to the Fermi level are not treated accurately. Treating those states separately improves the accuracy significantly (second column). 
   }
    \label{fig:cheb-err}
\end{figure}

In each iteration step towards a converged HF solution, we employ either exact diagonalization (ED) or Chebyshev expansion {of the local density} (as described below) to evaluate the thermal expectation values $\rho_i$ and $m_i^z$. {Thermal expectation values are computed by summing contributions from HF orbitals, weighted according to a thermal Fermi-Dirac distribution}. Those expectation values are then fed back into the HF Hamiltonian for the next iteration step. In each of the self-consistent Hartree-Fock (SCHF) iteration steps we tune the chemical potential such that the system is at half-filling. 
For Figs.~1-4 of the main text, we used ED as it is straightforward to implement and numerically exact. For Fig.~5 of the main text, we used Chebyshev expansion to evaluate the local density as it allows us to access unprecedentedly large systems. 

\subsection{Direct Chebyshev expansion for SCHF}

In the conventional kernel polynomial method (KPM) \cite{weibe2006_kpm}, damping kernels, e.g. Jackson or Lorentzian kernels, are applied to suppress so-called Gibbs oscillations in the Chebyshev expansion of physical quantities. However, we found that higher accuracy can be achieved with a direct Chebyshev expansion (without kernels) in evaluating $\rho_i$ and $m_i$. At finite $T$, the Fermi-Dirac distribution is smooth in energy and machine precision can be achieved with a relatively small number $N_\mathcal{C}$ of expansion terms. 
Also at $T=0$, the direct Chebyshev expansion turns out to be more accurate than damped KPM at the same $N_\mathcal{C}$, as the damping kernels mainly introduce extra sources of inaccuracy.

\subsection{Evaluating local densities}
The most time-consuming step in the above Hartree-Fock procedure lies in the evaluation of the local electron density. In exact diagionalization, it can be directly evaluated with 
\begin{equation}
    n_{i\sigma} = \sum_m f(E_m) \left|\psi_{m}^{i\sigma}\right|^2\,,
\end{equation}
where $E_m$ and $\psi_m$ are the eigenenergy and eigenstate of the Hartree-Fock Hamiltonian, respectively, and $f(E)$ is the finite temperature Fermi-Dirac distribution. 
{Chebyshev expansion 
instead uses the fact that the local density } can be expressed as
\begin{equation}
    n_{i\sigma} = \left[f(H)\right]_{i\sigma, i\sigma}\,.
\end{equation}
{Here $H$ is the HF Hamiltonian with chemical potential $\mu$, adjusted to ensure half-filling. }We can then linearly rescale the HF Hamiltonian to $\tilde{H} = \alpha H + \beta$, with $\alpha,\beta$ chosen  such that its spectrum lies inside the interval $[-1, 1]$. {One can then expand the function $f(H)=f((\tilde{H}-\beta)/\alpha))$ into Chebyshev polynomials $T_m(\tilde H)$, whose matrix elements can be efficiently evaluated  numerically. This results in the series} 
\begin{equation}
    n_{i\sigma} 
    \approx 
    c_0 \mu_0^{i\sigma}
    + 
    2 \sum_{m=1}^{N_\mathcal{C}-1} 
    c_m \mu_m^{i\sigma},
\end{equation}
where the expansion is truncated at $m= N_{\cal C}$.
$\mu_m^{i\sigma} = \braket{i\sigma|T_m(\tilde{H})|i\sigma}$ is the $m$-th Chebyshev moment of the local density and $c_m$ is the Chebyshev coefficient of the Fermi-Dirac distribution,
\begin{equation}
    c_m = \frac{1}{\pi} \int_0^{\pi} f\left(\frac{\cos\theta - \beta}{\alpha}\right) \cos(m\theta) d\theta\,. 
\end{equation}
As we mentioned, in previous works, damping kernels, e.g. the Jackson or Lorentzian kernels, were introduced {to tame the sum over $m$} and suppress Gibbs oscillations. However, the evaluation of $n_{i\sigma}$ involves integration over the spectrum and Gibbs oscillations have only a weak effect on the integral. We found it to be more efficient and accurate to calculate $n_{i\sigma}$ without such damping kernels, which appeared to only introduce extra errors.

Fig.~\ref{fig:cheb-err} shows the error of the Chebyshev expansion (compared to numerically exact ED) without using damping kernels. Machine precision can be achieved for finite temperature {when $N_\mathcal{C} \approx O(30 t/T)$ is chosen to be large enough.} 
For strictly $T=0$, the Chebyshev expansion converges much more slowly than at finite $T$ due to the discontinuity in the Fermi-Dirac distribution. Higher accuracy can be achieved by correcting the Chebyshev expansion by using exact eigenstates near the Fermi energy{, correcting a low, but finite $T$ Chebyshev expansion with the contributions from eigenstates in the thermally relevant window, obtained accurately {and efficiently with the Chebyshev filter method \cite{Pieper2016, Zhang2025}} (second column in Fig.~\ref{fig:cheb-err}). 

The evaluation of the Chebyshev moments $\mu_{m}^{i\sigma}$ can be carried out very efficiently on modern GPUs with high bandwidth memory. It allows one to study large system sizes that were inaccessible in previous work, using only a few ($\le 4$) GPUs.

\section{More Results}\label{sec:results}

\subsection{Correlation between maximum of conductivity and formation of local moments}
Fig.~\ref{fig:sigmaxx-Siz-T} shows the temperature dependence of $\sigma_{xx}$ and the local moment density $\langle \left| S_i^z \right|\rangle$ at different dopant densities. Local magnetic moments form at sufficiently low temperatures (blue curves in Fig.~\ref{fig:sigmaxx-Siz-T}). As the dopant density decreases ($R$ increases), the local moments start to form at higher temperature and reach a larger density in the low $T$ limit. While local moments are still sparse or completely absent at higher temperature the conductivity $\sigma_{xx}$ increases upon decreasing the temperature. That trend is eventually cut off by the formation of a sufficient density of local moments, $\sigma_{xx}$ reaching a maximum at the temperature $T^*_\sigma$ [Fig. 4(c) in the main text]. $T^*_\sigma$  increases with increasing $R$, and lies not far below the temperature $T^*_s$ where local-moments start forming. This is interpreted physically in the main text.

\begin{figure}
    \centering
    \includegraphics[width=0.99\linewidth]{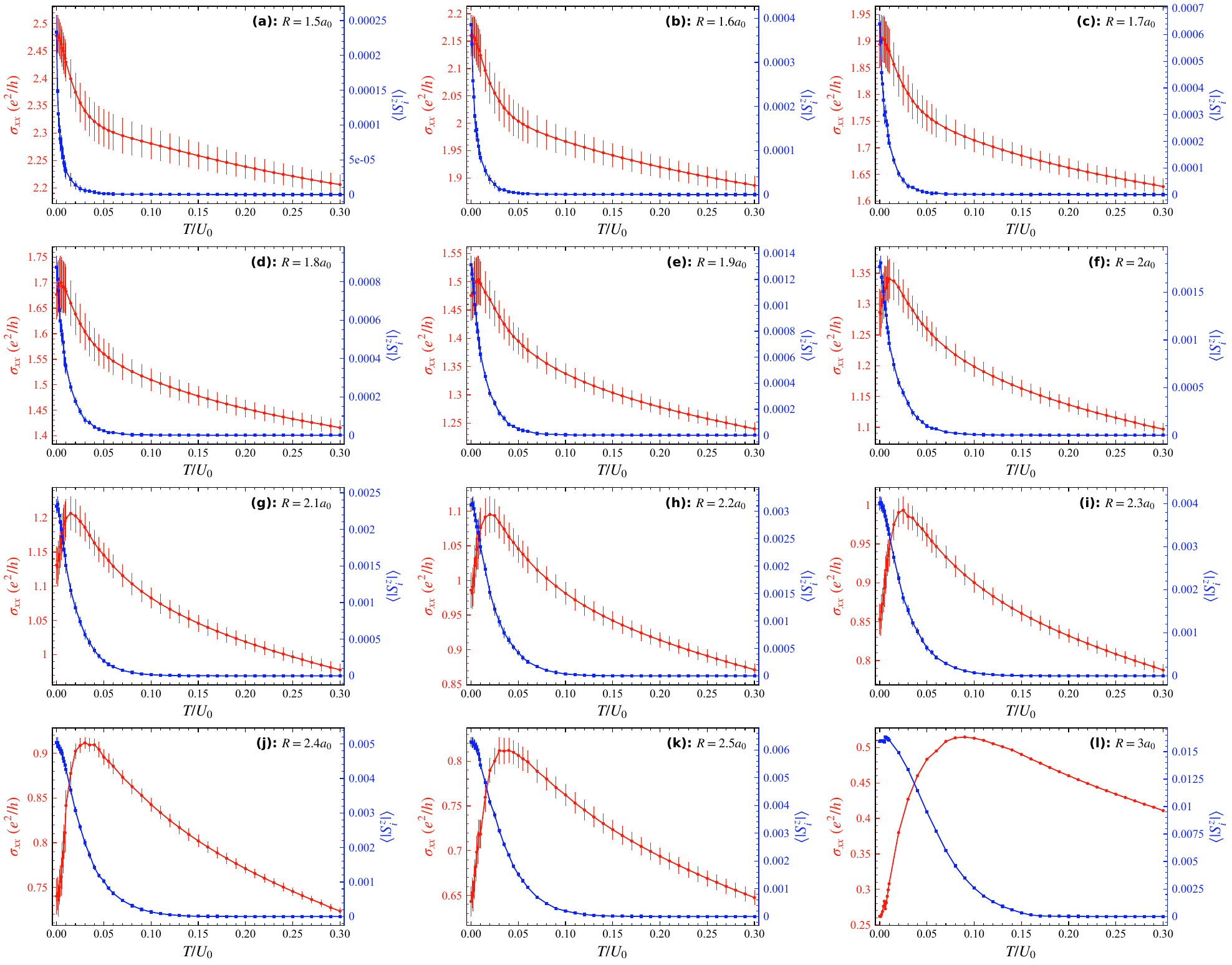}
    \caption{The $T$-dependence of the average conductivity $\sigma_{xx}$ and the local moment density $\left<\left|S_i^z\right|\right>$ for various mean dopant spacings $R$. 
    At all densities, a downturn of conductivity occurs at a temperature slightly below the onset of local moments. 
    The data are averaged over $10$ disorder realizations, except for $R=3a_0$ where a single realization was used.}
    \label{fig:sigmaxx-Siz-T}
\end{figure}

\subsection{Multiple HF solutions at low temperature}

At high temperatures the self-consistent Hartree-Fock (SCHF) equations have a unique solution. However, at low enough temperatures, and certainly at temperatures below $T^*_s$ where local moments form, multiple solutions with similar energies exist. The solution we obtain depends on the  charge and spin configurations assumed at the initial stage of the convergence procedure. (Note that to allow for local moments to emerge, we need to break the time reversal symmetry in the initial condition, by assuming a small magnetization.) Fig.~\ref{fig:charge-spin-rounds} shows the charge and spin configurations of the SCHF states obtained from different initial conditions. Fig.~\ref{fig:charge-spin-rounds} (a) and (c) show the charge and spin polarization maps of one converged state, which we use as reference state; its charge and spin configurations are denoted by $n_i^{(0)}$ and $(S_i^z)^{(0)}$, respectively. We then flip a randomly selected fraction $r\in(0, 1)$ of the spin polarizations $(S_i^z)^{(0)}$, and use the resulting configuration as the initial state for a new SCHF calculation. The bottom six rows show the charge and spin configurations of the newly SCHF converged states. The second and the fourth columns show the differences in the charge and spin configurations relative to the reference state. The converged states clearly exhibit different charge and spin configurations. However, the differences between the charge distributions are much smaller than those between the spin configurations. The local magnetic moments reside at nearly the same locations in all states, while their magnitude and polarization directions may differ. 

Our results suggest that the Hartree-Fock energy landscape exhibits several local minima  differing in their magnetic texture, possibly giving rise to spin-glass-type order in the low-$T$ limit. In contrast, the charge sector shows no analogous glassy structure. This  feature may distinguish $\delta$-layers from the related 2-DEGs in MOSFETs. Indeed $\delta$-layers have a much higher density, which, together with the small ratio between nearest neighbor and onsite Coulomb repulsion (which we neglected here entirely), renders frustration due to long-range Coulomb interactions unimportant. In MOSFETs instead the latter are much stronger and presumably entail strong frustration in the charge sector and possibly an electron glass.  

\begin{figure}
    \centering
    \includegraphics[width=0.8\linewidth]{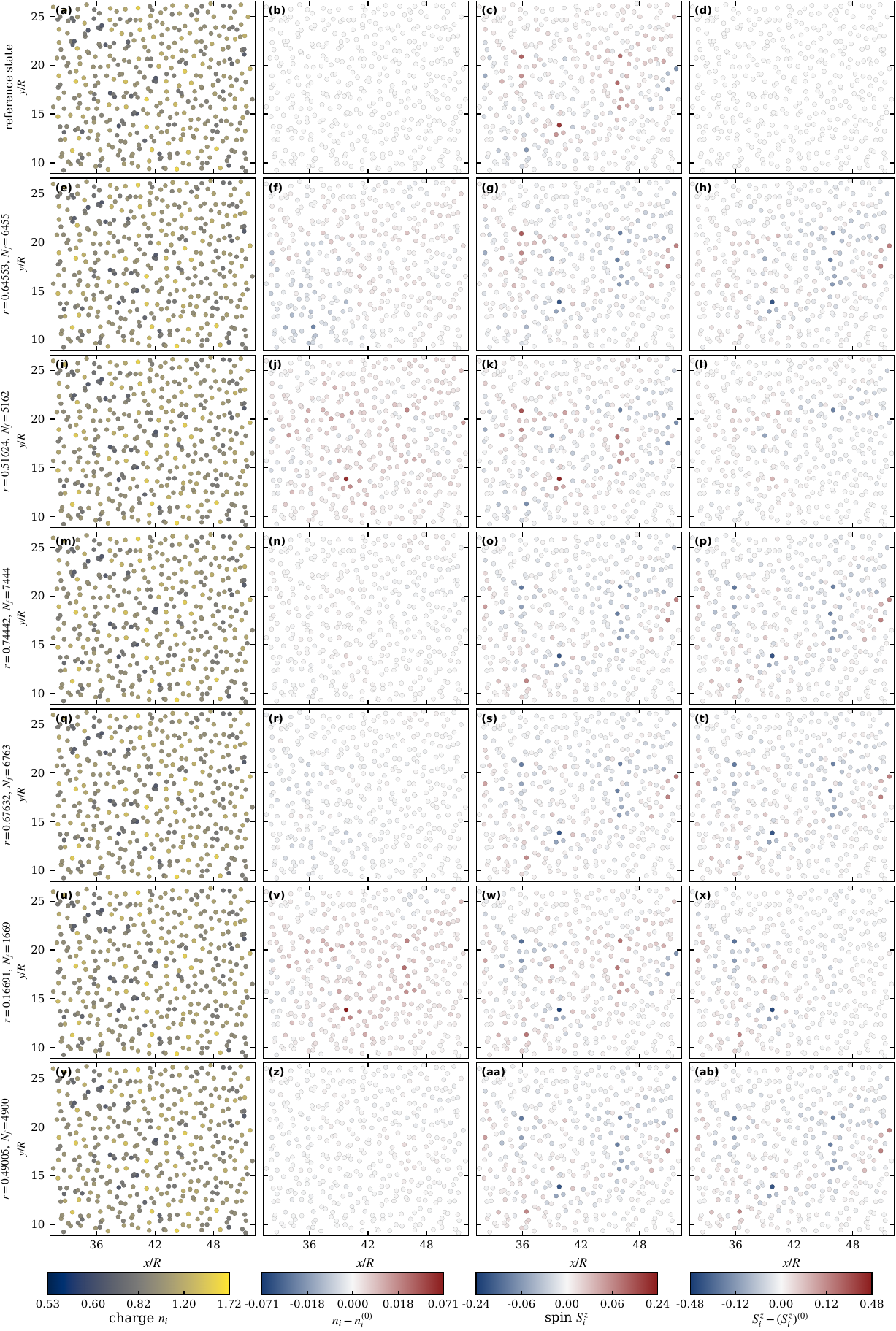}
    \caption{The charge and spin texture of 7 self-consistent Hartree-Fock solutions, converged from different initial conditions at $T=0$, for the same dopant configuration. A reference state (first and third column) is obtained from a randomly chosen initial state. We then flip a random fraction $r\in (0, 1)$ (indicated on the left) of the magnetizations in the reference state and feed it back as the initial state for new rounds of Hartree-Fock iterations. The plots show a $20\times 20$ patch selected from a $100\times 100$ lattice to display the variations from state to state. The variations in electron density are typically at most at the level of a few percent, while the local moments often flip entirely ($\Delta S^z_i = \pm O(1)$) between two different states. 
    }
    \label{fig:charge-spin-rounds}
\end{figure}

\bibliography{ref}